\documentclass[sigconf]{acmart}
\usepackage{epsfig}
\usepackage{amssymb}
\usepackage{amsmath}

\usepackage{amsfonts}
\usepackage[utf8]{inputenc}
\usepackage{caption}
\usepackage{graphicx}
\usepackage{textcomp}
\usepackage{booktabs}
\usepackage{algorithm}
\usepackage{algorithmic}
\usepackage{tikz}
\usepackage{adjustbox}
\usepackage{tabularx,multirow,booktabs,blindtext}
\usepackage{mathdots}
\usepackage{amsmath}
\usepackage{tikz}
\usepackage{mathdots}
\usepackage{yhmath}
\usepackage{cancel}
\usepackage{color}
\usepackage{pgfplots}
\pgfplotsset{width=8.2cm,compat=1.9}
\usepackage{siunitx}
\usepackage{array}
\usepackage{multirow}
\usepackage{amssymb}
\usepackage{gensymb}
\usepackage{tabularx}
\usepackage{booktabs}
\usetikzlibrary{fadings}
\usepackage{yhmath}
\usepackage{cancel}
\usepackage{siunitx}
\usepackage{array}
\usepackage{multirow}
\usepackage{amssymb}
\usepackage{gensymb}
\usepackage{tabularx}
\usepackage{booktabs}
\usetikzlibrary{fadings}
\usepackage{latexsym} 
\pgfplotsset{compat=1.11,
        /pgfplots/ybar legend/.style={
        /pgfplots/legend image code/.code={%
        \draw[##1,/tikz/.cd,bar width=3pt,yshift=-0.2em,bar shift=0pt]
                plot coordinates {(0cm,0.8em)};},
},
}
\copyrightyear{2019}
\acmYear{2019}
\setcopyright{acmlicensed}
\acmConference[CARS- 2019]{CARS RecSys19: Context-Aware Recommender Systems Workshop }{20 September 2019}{Copenhagen, Denmark}
\acmBooktitle{CARS- 2019: Context-Aware Recommender Systems Workshop, September 20, 2019, Copenhagen, Denmark}
\acmPrice{}
\acmDOI{}
\acmISBN{}

\begin{document}

%
\title[On-device User Intent Prediction for Context Aware Recommendation]{On-device User Intent Prediction for Context and Sequence Aware Recommendation}

%
\author{Benu Madhab Changmai\qquad Divija Nagaraju\qquad Debi Prasanna Mohanty\qquad Kriti Singh\qquad Kunal Bansal\qquad Sukumar Moharana}

\affiliation{Samsung Research Institute, Bangalore \\
  \{\tt b.changmai, d.nagaraju, debi.m, kriti.18892, kunal.bansal, msukumar\}@samsung.com}

%
\renewcommand{\shortauthors}{Benu and Divija, et al.}

%
\begin{abstract}
The pursuit of improved accuracy in recommender systems has led to the incorporation of user context. Context-aware recommender systems typically handle large amounts of data which must be uploaded and stored on the cloud, putting the user's personal information at risk. While there have been previous studies on privacy-sensitive and context-aware recommender systems, there has not been a full-fledged system deployed in an isolated mobile environment. We propose a secure and efficient on-device mechanism to predict a user's next intention. The knowledge of the user's real-time intention can help recommender systems to provide more relevant recommendations at the right moment. Our proposed algorithm is both context and sequence aware. We embed user intentions as weighted nodes in an n-dimensional vector space where each dimension represents a specific user context factor. Through a neighborhood searching method followed by a sequence matching algorithm, we search for the most relevant node to make the prediction. An evaluation of our methodology was done on a diverse real-world dataset where it was able to address practical scenarios like behavior drifts and sequential patterns efficiently and robustly. Our system also outperformed most of the state-of-the-art methods when evaluated for a similar problem domain on standard datasets.
\end{abstract}

%
%


%

\maketitle

\section{Introduction}
Mobile devices are a crucial part of our lives today.  Recent studies done in the United States\footnote{https://www.emarketer.com/corporate/coverage/be-prepared-mobile} show that on an average, users spend over four hours of their time on their mobile devices daily.  This increase in activity has led to an emphasis on curated information to decrease user's search time, prompting the development of recommendation systems on most digital platforms which can serve the user with content and services at the right time and place.  These systems range from predicting the next app a user will use \cite{NextApp} to recommending the location of the next check-in a user will post on social media \cite{location-based-POI} . The past year has shown an unprecedented  scrutiny  surrounding  the  issue  of  user  data  privacy. Studies \cite{ThirdPartyTracking} show that 1 in 5 applications share their data with more than 20 third parties. The judicious use of data by third parties cannot be ensured but an attempt can be made to provide useful on-device recommendations in an efficient manner without data leaving the user's device. This leads us to addressing the following difficult but important research questions:
\begin{itemize}
    \item \textbf{RQ1} Can there be a method of profiling user behavior and patterns which can be run entirely on a mobile device without any external server or cloud based intervention?
    \item \textbf{RQ2} Are individual sequential choices good enough data points to provide recommendations and can they completely replace collaborative methods of making and mapping user profiles? 
    \item \textbf{RQ3} Can there be a method that can adapt to ever changing user behavior patterns?
\end{itemize}

In this paper, we present an on-device intent prediction methodology called WIME that predicts the user intent based on user’s previous action sequences and contextual information. This is done by storing a user’s intents in the form of as nodes embedded into an n-dimensional vector space. The number of dimensions, $\displaystyle{n}$ corresponds to the number of contextual parameters taken into consideration. The proposed method also accommodates a user’s behavior drift \cite{interestDriftPOI} by maintaining and updating node weights with respect to regularity and recency of usage. Recommendation of next user action is done based on a nearest neighbor search \cite{kdTree} followed by a sequence matching method \cite{winkler}. Our algorithm also addresses the solution to cold start problem, eliminating the necessity of devising a collaborative approach. The proposed algorithm is proven to be computationally lightweight and memory efficient enabling it to run on personal devices locally in the interest of protecting user data. For demonstration purposes, we have modelled intents as a set of actions a user performs on a mobile device daily such as listening to music, booking a cab, reading the news, etc. The main contributions of this paper are as follows:
\begin{itemize}
    \item We  propose  a  novel  method  (WIME)  for  modelling user's  intents  as  weighted  nodes  in  a  multidimensional vector space in the form of embeddings.
    \item We develop an on-device prediction engine, which enables the device to predict user intents in real time without requiring any server intervention and thus preserving the user privacy.
    \item We address the dynamic user behavior patterns as well as context dependent user intents, without involving any complex or computationally expensive deep learning mechanisms.
    \item We  evaluate  the  proposed  method  by  detailed  experimentation on real world datasets and demonstrate that our method outperforms the state of the art methods.
\end{itemize}
The paper is sectioned as follows: Section 2 describes the related work, Section 3 and 4 mention the datasets the methodology was tested on and a detailed problem description respectively. Section 5 explains the methodology, 6 states the experimental results and observations. Finally, Section 7 describes the real world applications of WIME.

\section{Related Work}
Previous literature was studied keeping the following aspects in mind to ensure the goal of the study is fulfilled.
\begin{itemize}
    \item Can the approach be modified to fit the resource constraints of a mobile device?
    \item Whether the approach ensures optimal utilization of the data available on mobile devices.     
    \item How does the approach counteract the problem of cold start?
    \item Whether the approaches are resistant to behavior drifts in usage patterns.     
    \item Which aspects of the approach make assumptions about the user and can be replaced with an appropriate feedback mechanism so as to adopt a more user centred methodology.
\end{itemize}
This research discusses three different types of recommender systems namely, context aware, sequence based and privacy centred. 

\subsection{Context Aware Recommender Systems}
In a recommender systems setting, context refers to the set of factors that affect the choices a user makes. These are usually characterized as spatial, temporal, social and activity based. The emergence of using such factors in a recommender system began in 2005 through the study\cite{CARSEmergence}. Incorporation of such factors into a recommender system helps improve its accuracy to a large extent. When developing a recommender system for a mobile scenario, the user context is easily accessible. This has been used previously in \cite{CARSMobileSassi} and several others as mentioned in \cite{CARSMobile2}. However, most of the approaches are not in an isolated mobile environment and hence raise privacy concerns. Furthermore, most of these systems do not incorporate the sequence of user choices into the recommender system thus providing scope for improvement. 

\subsection{Sequence based recommender systems}
Another approach that is often used in recommender systems is considering the sequence of actions that a user has undertaken.  This is used in systems where there are recurrences in the recommendations such as in Point Of Interest (POI) recommendation where a user may visit the same place multiple times or in online streaming services where there is a definite sequence present in what a user watches.  A detailed profiling of such systems is done in \cite{SeqAwareSurvey}. The primary methods for building such systems include usage of Markov chains \cite{MC} and more recently Factorizing Personalized Markov Chains \cite{FPMC}, which combines the matrix factorization method, mentioned above with Markov chains. These systems fall short of utilizing the complete available information since they do not consider contextual data discussed previously.

\subsection{Privacy centred recommender systems}
The dearth of privacy centred or even privacy aware recommender systems is well known as clearly stated in \cite{RecsysPrivacyTutorial}. While incorporating privacy in recommender systems can be done in numerous algorithmic techniques such as homomorphic encryption, obfuscation and differential privacy, we sought to take the simpler path by making modifications in the architecture itself. \cite{clientBasedRecSys} shows that client side personalisation or on-device personalisation greatly improves the user's perception of privacy thus validating our claim. The approach taken here is to keep the methodology computationally simple to facilitate storage and running of the algorithm completely on-device.

Our proposed approach focuses on combining the capability of sequential and context aware recommender systems and doing so completely on-device. 

\section{Dataset}
For demonstration of varied user behavior, we prepared a real world dataset that records the daily activities of a user. We collected data of 54 volunteers with diverse lifestyles. We extracted their application logs, app notifications, GPS logs etc. from their smartphones for 21 weeks. Each activity was mapped to a particular intent of the user. We identified 113 unique intents from the dataset over 600 unique applications. This number can be increased or decreased for more granular or generic recommendations respectively. These intents, with their corresponding sequential information, temporal, spatial, and other context factors formed our dataset. Table \ref{table:dataset} illustrates a typical snapshot of the dataset we prepared. To compare the performance of our algorithm with other relevant state-of-the-art we also ran our algorithm on the Foursquare \cite{foursquare} dataset for New York and Los Angeles. The statistics of the dataset are summarized in Table 2.

\begin{table}
\begin{tabular}{|>{\centering\arraybackslash}p{0.017\textwidth}|>{\centering\arraybackslash}p{0.066\textwidth}|>{\centering\arraybackslash}p{0.04\textwidth}|>{\centering\arraybackslash}p{0.042\textwidth}|>{\centering\arraybackslash}p{0.042\textwidth}|>{\centering\arraybackslash}p{0.060\textwidth}|>{\centering\arraybackslash}p{0.065\textwidth}|>{\centering\arraybackslash}p{0.07\textwidth}|}
\hline 
 ID & Intent & \multicolumn{3}{l|}{Day::Time} & Latitude & Longitude & Preceding seq \\
\hline 
 3 & Read News & \multicolumn{3}{l|}{Mon::08:14} & 12.970  & 77.692 & -- \\
\hline 
 6 & Social Connect & \multicolumn{3}{l|}{Mon::08:32} & 12.970 & 77.692 & 3 \\
\hline 
 7 & Call Contact & \multicolumn{3}{l|}{Mon::08:47} & 12.970 & 77.692 & 6-3 \\
\hline 
 9 & Check Mail & \multicolumn{3}{l|}{Mon::10:05} & 12.980 & 77.697 & 7-6 \\
\hline 
 7 & Call Contact & \multicolumn{3}{l|}{Mon::10:11} & 12.980 & 77.697 & 9-7-6 \\
 \hline
\end{tabular}
\caption{A snippet of our real world dataset showing the contextual features and user's preceding action sequence.}
\label{table:dataset}
\end{table}
\begin{table}[h!]      
\begin{tabular}{|p{0.03\textwidth}|p{0.05\textwidth}|p{0.09\textwidth}|p{0.09\textwidth}|p{0.08\textwidth}|}
\hline 
  & \#User & \#Check-in &  \#Avg Check-in & \#Category \\
\hline 
 LA & 1445 & 91414 & 63.20 & 293 \\
\hline 
 NYC & 2579 \ \  & 157404  &  61.03 &  252 \\
 \hline
\end{tabular}
\caption{Foursquare Dataset Statistics}
\end{table}

\section{Problem Description}
For a user $\displaystyle u$, let $\displaystyle I=\{i_{1} ,i_{2} ,i_{3} ,i_{4} ...\}$ be the set of identified intent categories reflected by his/her actions until instance $\displaystyle t$ . Each intent $\displaystyle I_{X_{t}}$ occurs in a set of context $\displaystyle X_{t} =\left\{x^{1}_{t} ,x^{2}_{t} ,x^{3}_{t} ,...\right\}$where $\displaystyle x^{i}_{t}$ \ represents the metric of the $\displaystyle i^{th}$ \ context factor at instance $\displaystyle t$. For a given instance $\displaystyle t-1$, with a sequence of intents $\displaystyle I_{X_{t-n}} ...I_{X_{t-3}} I_{X_{t-2}} I_{X_{t-1}}$ observed till then, our goal is to predict the next intent $\displaystyle I_{X_{t}}$ that the user will express for the current real-time context $\displaystyle X_{t}$ at instance $\displaystyle t$. This intent can be used to recommend relevant content to the user.   

\section{Methodology} 
In this section, we describe a context aware weighted intent embedding generation process and its analysis considering a prior sequence of user intents to predict the next user intent to facilitate the recommendation process.
\subsection{Embedding Contextual Information}
The mobile environment the system has been built in was leveraged by taking additional factors into consideration and representing them in the appropriate form.
\paragraph{Temporal Embedding:}
Time is often modelled in discrete categories or slots while considering temporal features as a factor for prediction. However, this leads to a loss of granularity. In this work, time has been represented as a cyclic function of the user's current time. The features extracted thus are time of the day and time of the week to capture daily and weekly patterns respectively.\\ \indent
Time of the day is embedded as minutes past 00:00 am. The $\displaystyle \sin( x)$ and $\displaystyle \cos( x)$ values of the time are plotted, where, $\displaystyle x$ is the ratio of minutes past 00:00 am and total minutes in a day. This gives an embedding of 2 dimensions for time of the day. Time of the week is required to capture differences between the same time of different days. This has been embedded in a similar fashion as the time of the day and $\displaystyle x$ is the ratio of minutes past 00:00 Sunday midnight and total minutes in a week. Overall, we express time over 4 dimensions. The length of this embedding can be varied based on the amount of granularity the use case requires. \\ \indent Fig. 1 shows how time is plotted as a cyclic function. Every point in the curve represents a point of time.

\begin{figure}
 \begin{tikzpicture}[scale = 0.78] 
\begin{axis}[
    xlabel={sin(x)},
    ylabel={cos(x)},
    xmin=-1.1, xmax=1.1,
    ymin=-1.1, ymax=1.1,
    xtick={-1.0,-0.5,0.0,0.5,1.0},
    ytick={-1.0,-0.75,-0.5,-0.25,0.0,0.25,0.5,0.75,1.0},
    ymajorgrids=true,
    grid style=dashed,
]

\addplot[
    color=blue,
    mark=*,
    mark size=1pt
    ]
    coordinates {
(0.08353290843637881,0.9965050191585387)(0.2092630170368122,0.9778593915797152)(0.4660649667968004,0.8847504997029939)(0.7421708287796018,0.6702107585741959)(0.8995895416989381,0.43673636952501954)(0.9891123925568837,0.14716207014851757)(0.9995201707484881,0.030974639092542635)(0.9996852848389379,0.025086475967971488)(0.9869729770623776,-0.16088611670566075)(0.9847445476525166,-0.17400625238375916)(0.9480232558708657,-0.3182010470253094)(0.9217503888731543,-0.38778372917412207)(0.8988259231719091,-0.43830578348245114)(0.7121281624814388,-0.7020494855783384)(0.5680233120357954,-0.8230124646588808)(0.42946059995432534,-0.9030855956590554)(0.3827506177686793,-0.9238517005438126)(0.32701173532127703,-0.9450202775402056)(0.3109529103012239,-0.9504253193045727)(0.21366981543469438,-0.9769059371158022)(0.14154917068125747,-0.9899312260351465)(0.11197479963886095,-0.993711046655836)(0.0899083287242317,-0.9959500451458474)(-0.10698714613208128,-0.994260403798981)(-0.15435114373558245,-0.9880160547417828)(-0.22367541455210294,-0.9746636901644304)(-0.3885879603733062,-0.9214116327966094)(-0.6184655428482593,-0.7858119191698533)(-0.7761381269359211,-0.6305629293864333)(-0.8132700593520054,-0.5818864241083355)(-0.8594807669762105,-0.5111680850737701)(-0.9620197313291151,-0.2729799196524483)(-0.9992768085541044,0.03802446430817694)(-0.9989306027408966,0.04623473702432958)(-0.8986983871619978,0.4385672227976277)(-0.8978675500026393,0.44026567280706536)(-0.8542698711721574,0.5198297675273182)(-0.823218949352743,0.5677240187155781)(-0.8184005329065943,0.5746482121595807)(-0.8003391394443621,0.599547547633595)(-0.7689348265740154,0.6393271717059967)(-0.731353701619171,0.6819983600624979)(-0.5754811349186445,0.8178150544913861)(-0.4950164235916872,0.8688836172782264)(-0.4677369953754361,0.8838676955049094)(-0.4346417489368663,0.9006034366362934)(-0.32769888692989635,0.9447822180295875)(-0.2616277104629439,0.9651688666331493)(-0.2120355374097402,0.9772619561178889)(-0.006399496909477121,0.9999795230099993)};
\end{axis}
\end{tikzpicture}
\hfill
\captionof{figure}{ Plotting time $\displaystyle x$ as a cyclic value}
\end{figure}
\paragraph{Geo-spatial Embedding:}
Location of the user is another major factor that influences his/her intents. We consider the latitude and longitude in our dataset as the metric for location. The location coordinates have been scaled to a suitable unit such that the variations in location of any order can be captured.

\subsection{Capturing Sequence}
An important factor while predicting the user intent at an instance is to consider the sequence of actions the user has taken before that point. As it is observed from our real world dataset, there may be variations in the sequence of actions for the same user state. 
Consider a scenario where the user follows sequence A on most days at a particular state:  
A: \textbf{Check Mail }$\displaystyle \rightarrow $\textbf { Read News }$\displaystyle \rightarrow $\textbf{ Commutes to Office } $\displaystyle \rightarrow $\textbf{ Listen Music }\\ 
While on certain days, he/she follows the sequence B at the same state:
B: \textbf{Check Mail }$\displaystyle \rightarrow $\textbf{ Attend Calls }$\displaystyle \rightarrow $\textbf{ Commutes to Office } $\displaystyle \rightarrow $\textbf{ Read News } \\ 
It can be inferred that the user reads news during commute only if he/she skips it before because of some other action. This shows that the user's intent may change despite his/her real-time context remaining the same. This implies that the previous sequence of actions is also a contributing factor to his/her current intent. We will discuss how to leverage this sequence information in Section 5.6.
\subsection{Node Formation}
\begin{figure}[h!]
\caption{As the user performs daily activities along the course of the day, we plot these activities as intention nodes. For representation purposes, only 2 dimensions (location and time) are shown.}
\includegraphics[scale=0.37]{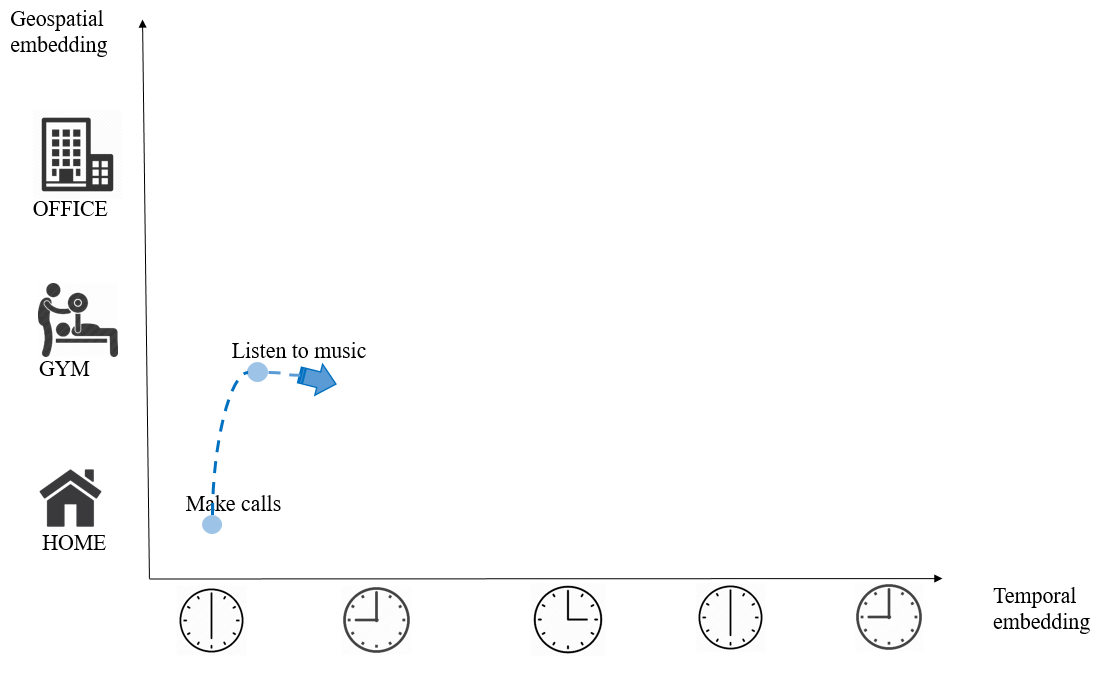}
\label{fig:nodeformation1}
\end{figure}

\begin{figure}[h!]
\caption{The nodes get accumulated in the vector space and gain weight based on the usage frequency.}
\includegraphics[scale=0.37]{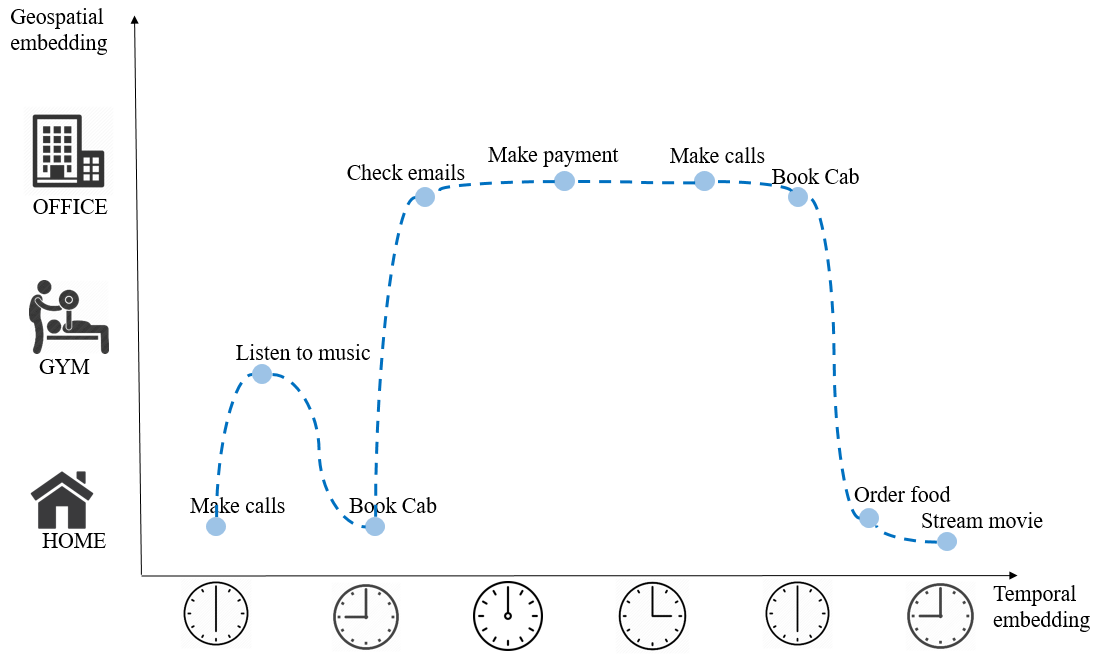}
\label{fig:nodeformation2}
\end{figure}

Each intent identified with a particular set of context factors leads to either (1) Node creation: Marked as a new node, (2) Node fusion: Merged with a pre-existing node of the same intent.\\ \indent
In order to take one of the above steps, the neighborhood of the point in the vector space is explored. If there already exists a node of the same intent, then it is merged with the pre-existing node instead of forming a new node. Every time such node fusion occurs, weight of the node is incremented. This process is illustrated in Figure \ref{fig:nodeformation1} and Figure \ref{fig:nodeformation2}.\\ \indent
We utilize the k-dimensional tree (k-d tree)\cite{kdTree} approach to speed up the search process at the cost of an overhead for maintaining the k-d tree. The time complexity of search is reduced from  $\displaystyle O(n)$ to $\displaystyle O(\log n)$ while adding an overhead of $\displaystyle O(\log n)$ during node insertion where $\displaystyle n$ is the total number of nodes in the tree. Therefore, search time complexity is optimized by  $\displaystyle O(n)$ – $\displaystyle O(\log n)$, which is significant when it comes to real-time predictions on-device. 
\subsection{Node Weight}
Each node in the vector space represents a dynamic intent of the user, which is sensitive to changes in the user’s preferences, interests and habits. To capture the dynamic relevance of the node, we took into consideration the frequency and decay of the node. We formulated the weight of the node as follows:\begin{equation}
w^{new}_{i} = k^{d} \ w^{old}_{i} +1
\end{equation}
where,$\displaystyle w^{new}_{i}$ is the updated weight of the $\displaystyle \ i^{th}$ node, $\displaystyle d$ is the number of days (daily pattern) or weeks (weekly pattern) since the node was last traversed and $\displaystyle k$ is the decay factor ($\displaystyle 0.4 < k < 1$). It can be pointed that when k assumes the value of 1, Eq. 1 becomes:
$\displaystyle w^{new}_{i} = w^{old}_{i}+1$   

In other words, this node weight gets simplified to a frequency driven metric. Decay factor is a necessary addition to the system since it eliminates redundant nodes thereby reducing memory usage and search time complexity. The nodes in the neighborhood of the current node whose weights have gone below a predefined threshold (For our case, 0.3) are considered obsolete and are removed from the vector space. Decay factor is empirically chosen to provide a balance between recency and regularity.

\subsection{Drifting Average}
Due to the user’s dynamic behavior, there are some nodes whose occurrence pattern may change over time gradually, which is termed here as gradual drift or simply drift. However, in absence of a node updating method, the system will not be able to cope with the changing user actions. \\
For example, a user initially leaves for work at around 09:00 am. For ease of understanding, the geospatial coordinates of the user's home is written as home. Therefore, the node representing his/her commute to office is: \\ \indent
node 1:\{Commutes to office, 09:00 am, home\} \\
However, his/her routine gradually shifts along the course of time due to change in work schedule. It is observed that this shift is gradual and happens over multiple iterations. Due to the large distance along the time dimension in the vector space from node 1, this intent would be wrongly perceived as a new one once it crosses the threshold distance from node 1. As a result, an extra node would be created. To avoid this, we introduce the drifting average method so that the position of the nodes in the vector space can be regularly updated, reflecting the user’s most recent behavior.
Every time a node undergoes fusion, the features of the existing node are adjusted according to the new addition. The feature value $\displaystyle f_{i}$ of $\displaystyle node_{i}$ is updated as follows:
\begin{equation}
 f_{i j}^{new} =\frac {f_{i j}^{old} \ *\ w_{i} \ +\ f_{i j}^{*}}{w_{i} +1}\\
\end{equation}where $\displaystyle f_{i j}^{new}$ and $\displaystyle f_{i j}^{old}$ are the new and old $\displaystyle j^{th}$ feature value of  $\displaystyle node_{i}$ while $\displaystyle f_{i j}^{*}$ is the realtime value of the $\displaystyle j^{th}$ feature of the intent to be merged with $\displaystyle node_{i}$.
This way the nodes will gradually keep shifting in the vector space when the user drifts along any dimension. For the example we posed earlier, we arrive at the following configuration of the node by following the drifting average method:\\ \indent
node 1:\{Commutes to office, 09:30 am, home\}
\subsection{Sequence Comparison}
 Next we discuss how intent sequences captured in Section 5.2 are used to make sequence-based predictions. It is important to highlight the fact that we leverage past sequence to predict the next action at the current context. While most sequence-based systems use sequence to predict the next action in the sequence, which may not necessarily be at the current context. We compare the immediate past sequence from the point of context for which the next intent is to be predicted with the sequence information previously stored in the nodes to find the node with the most matching sequence. There are two important aspects to look into while incorporating sequence comparison into the prediction mechanism:
\begin{itemize}
\item Determining sequence length
\item Determining a sequence comparison metric
\end{itemize}
Here, we safely assume that intents beyond a time limit in the immediate past will not have any major influence in determining future intents. For our dataset, we set this time limit to 90 minutes. As we are limiting the sequences by time and not by length, there is a possibility that two sequences having different lengths need to be compared. 
For comparison of sequences, we evaluated Levenshtein distance \cite{levenshtein} as the distance metric to find similarity score of sequences of different length. The Levenshtein distance between two strings a,b of length $\displaystyle |a|$ and $\displaystyle |b|$ respectively is given by $\displaystyle lev_{a,b}( |a|,|b|) \ $  where
\begin{equation}
lev_{a,b}( i,j) = \begin{cases}
max( i,j),\ \ \ \ if\ min( i,j) =0,\\
min\begin{cases}
lev_{a,b}( i-1,j) +1 &  \\
lev_{a,b}( i,j-1) +1 \ \ \ else&  \\
lev_{a,b}( i-1,j-1) +1_{\left( a_{i} \neq b_{j}\right)} & 
\end{cases} & 
\end{cases}
\\
\\
\end{equation}
This distance becomes stringent while differentiating among sequences having similar composition but different order. Moreover, this metric outputs a discrete number as the distance thereby rendering it unsuitable for a normalized distance metric. \\ \indent
Another distance metric that overcomes the limitations of Levenshtein distance is Jaro distance. Jaro-Winkler \cite{winkler} is a modified version of this algorithm that supports the idea that similarity near the start of the string is more significant than similarity near the end of the string. Therefore, when we consider our intents to be in the order of decreasing recency, the Jaro-Winkler gives preference to more recent intents in the sequence. According to the paper \cite{jaroComparison}, Jaro-Winkler works faster than Levenshtein or any variant of it.
Jaro–Winkler similarity $\displaystyle sim_{w}$ is given as:\\
\begin{equation}
sim_{w} =sim_{j} +lp\left( 1-sim_{j}\right)
\end{equation}
where
\begin{itemize}
\item$\displaystyle sim_{j}$  is the Jaro similarity of sequences $\displaystyle s_{1}$ and $\displaystyle s_{2}$:
\begin{equation}
sim_{j} =\begin{cases}
0 & if\ m=0,\\
\frac{1}{3}\left(\frac{m}{|s_{1} |} +\frac{m}{|s_{2} |} +\frac{m-t}{|m|}\right) & otherwise
\end{cases}
\end{equation}
\item$\displaystyle l$ is the length of common prefix at the start of the sequence up to a maximum of four characters
\item$\displaystyle p$ is a constant scaling factor. The standard value for this constant is $\displaystyle p$ = 0.1
\end{itemize}
In our work, Jaro-Winkler distance has been used for sequence comparison. The next step is to determine how to incorporate the sequence similarity score as found in Eq. 4 into the final prediction results.

\subsection{Intent Prediction}
Once the nodes start forming, WIME is then capable of providing predictions. A given context is referred to as query position in this section. At a query position, a prediction is initiated. The prediction process can be enumerated in the following steps:\\

\begin{enumerate}
\item Given an input set of context features such as {time $\displaystyle t$, location $\displaystyle l$, etc.}, WIME identifies the position in the vector space pointed by the features and finds the nearest $\displaystyle{n}$ (5 for our dataset) nodes in the neighborhood.
\item If one or more nodes are found within the region, then they are ranked using their weight and the distance from the query position. Now we evaluate the prediction scores $\displaystyle{s_{i}}$ for each obtained node as follows:\\
\begin{equation}
score\left( s_{i}\right) =\tanh\left(\frac{w_{i}}{d}\right)\\
\end{equation}where $\displaystyle d$ is the Euclidean distance of\ $\displaystyle i^{th}$ node from query position.
\item The nodes are then filtered on a threshold basis so that only the nodes with scores above a cutoff (c) are evaluated further. This is done so that only the most relevant nodes move on to the next sequence comparison step as well as to minimize the number of comparisons.
\item The sequence information for the intent as discussed in Section 5.2 is quantified in this step. The recent action sequence until the current instance $\displaystyle{t}$ is compared with the corresponding past action sequences for the nodes to be ranked. For example, for a given spatio-temporal context say (Tuesday-08:00-Home), let the previous action sequence until that point be \textbf{Check Mail }$\displaystyle \rightarrow $\textbf{Read News }$\displaystyle \rightarrow $\textbf{Call Contact }. This sequence is now compared to the corresponding actions sequences of all the nodes found in step 1.
\item The node with the highest similarity score sequence is then chosen as the predicted intent for the given context.
This entire process is depicted in Figure \ref{fig:inference}.
\begin{figure}[h!]
\caption{The neighbourhood of the current user position (based on context) is searched and the intention of the nearest node is provided. }
\includegraphics[scale=0.37]{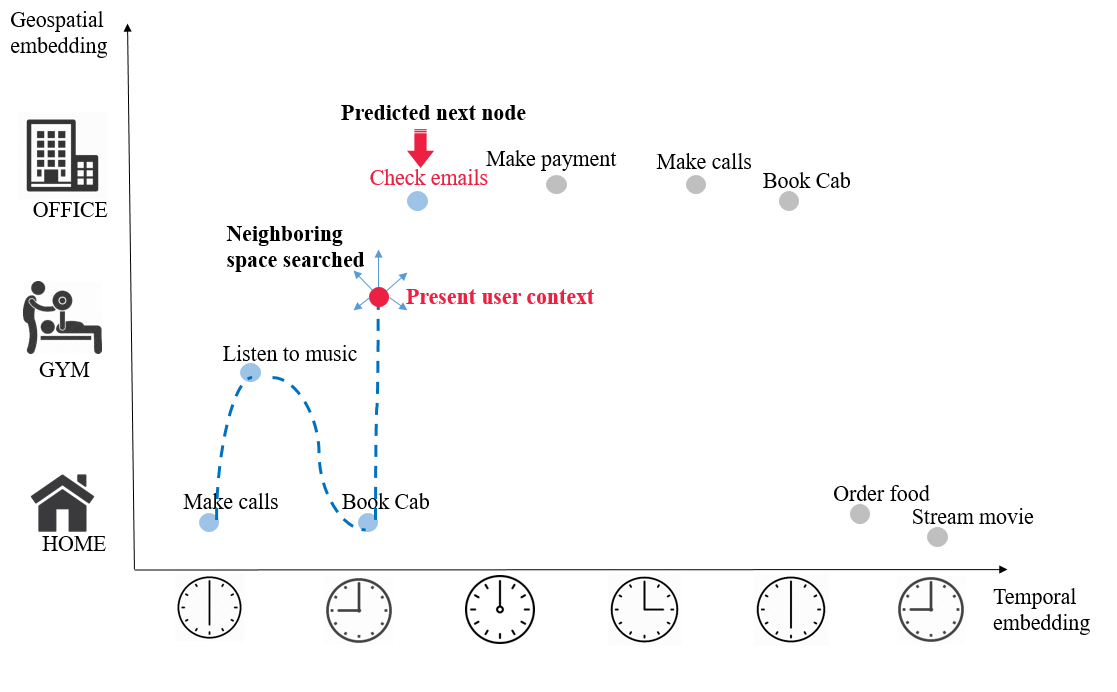}
\label{fig:inference}
\end{figure}
\end{enumerate}

\subsection{Providing Recommendations}
The contextual parameters given as input to and intention provided as output from WIME must be altered based on the type of recommendation required. In Section 6.1 we have noted the observations of WIME for next in-app recommendation i.e. it recommends not only the next app, but next function within an app that a user might be interested in. A description of the demo application is provided in Section 7.2. We have also tested the methodology for PoI recommendation in Section 6.2 to facilitate comparison with the current state of the art on standard datasets.     
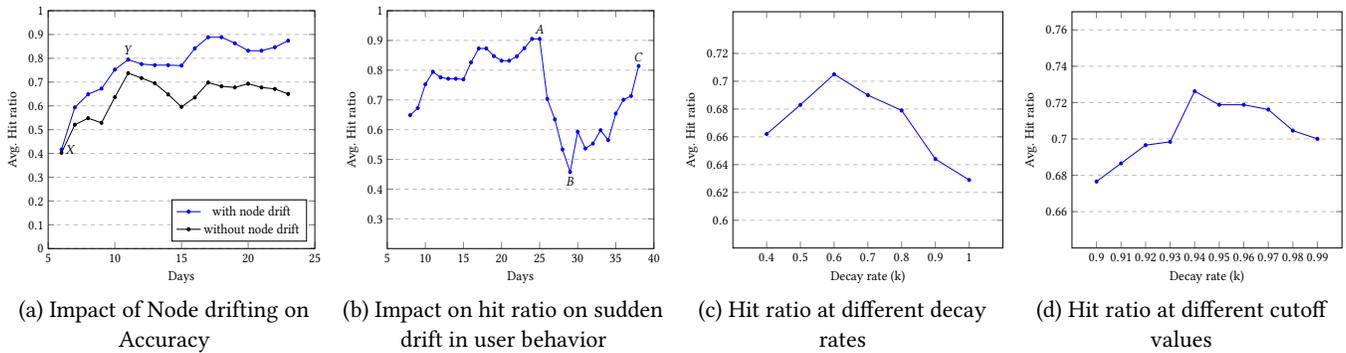
\begin{figure*}
\begin{minipage}[t]{.24\linewidth}\centering
\begin{adjustbox}{width=\linewidth,totalheight=1.5in}
 \begin{tikzpicture} 
\begin{axis}[
    xlabel={Days},
    ylabel={Avg. Hit ratio},
    xmin=5, xmax=25,
    ymin=0, ymax=1,
    xtick={},
    ytick={0,0.1,0.2,0.3,0.4,0.5,0.6,0.7,0.8,0.9,1},
    legend pos=south east,
    ymajorgrids=true,
    grid style=dashed,
]

\addplot[
    color=blue,
    mark=*,
    mark size=1pt
    ]
    coordinates {
( 6 , 0.41666666675 ) ( 7 , 0.593137255 ) ( 8 , 0.6486928105 ) ( 9 , 0.672385621 ) ( 10 , 0.752246732 ) ( 11 , 0.79452614375 ) ( 12 , 0.77573529425 ) ( 13 , 0.771139706 ) ( 14 , 0.771139706 ) ( 15 , 0.76905637275 ) ( 16 , 0.84166666675 ) ( 17 , 0.88854166675 ) ( 18 , 0.88854166675 ) ( 19 , 0.8630514705 ) ( 20 , 0.8318014705 ) ( 21 , 0.8318014705 ) ( 22 , 0.84638480375 ) ( 23 , 0.87395833325 )
    };\node [right] at (axis cs: 6 , 0.41666666675  ) {$X$};
\node [above] at (axis cs:   11 , 0.79452614375) {$Y$};
    \addlegendentry{with node drift}

\addplot[
    color=black,
    mark=*,
    mark size=1pt
    ]
    coordinates {
  ( 6 , 0.40277777775 ) ( 7 , 0.5204248365 ) ( 8 , 0.54820261425 ) ( 9 , 0.528594771 ) ( 10 , 0.63623366 ) ( 11 , 0.73733660125 ) ( 12 , 0.71691176475 ) ( 13 , 0.69485294125 ) ( 14 , 0.64797794125 ) ( 15 , 0.595894608 ) ( 16 , 0.63541666675 ) ( 17 , 0.69791666675 ) ( 18 , 0.68229166675 ) ( 19 , 0.677389706 ) ( 20 , 0.693014706 ) ( 21 , 0.677389706 ) ( 22 , 0.671139706 ) ( 23 , 0.65 )};

    \addlegendentry{without node drift}
   
\end{axis}
\end{tikzpicture}
\end{adjustbox}
\\(a) Impact of Node drifting on Accuracy
\end{minipage}
\hfill
\begin{minipage}[t]{.24\linewidth}\centering
\begin{adjustbox}{width=\linewidth,totalheight=1.5in}
 \begin{tikzpicture} 
\begin{axis}[
    xlabel={Days},
    ylabel={Avg. Hit ratio},
    xmin=5, xmax=40,
    ymin=0.2, ymax=1,
    xtick={},
    ytick={0.3,0.4,0.5,0.6,0.7,0.8,0.9,1},
    ymajorgrids=true,
    grid style=dashed,
]

\addplot[
    color=blue,
    mark=*,
    mark size=1pt
    ]
    coordinates {
 (8,0.6486928104999999)(9,0.6723856209999999)(10,0.7522467319999999)(11,0.79452614375)(12,0.77573529425)(13,0.771139706)(14,0.771139706)(15,0.76905637275)(16,0.82604166675)(17,0.87291666675)(18,0.87291666675)(19,0.8474264705)(20,0.8318014705)(21,0.8318014705)(22,0.84638480375)(23,0.87395833325)(24,0.90520833325)(25,0.90520833325)(26,0.703125)(27,0.634375)(28,0.53333333325)(29,0.45744047600000004)(30,0.59285714275)(31,0.53660714275)(32,0.5532738095)(33,0.59791666675)(34,0.56458333325)(35,0.6541666665)(36,0.700595238)(37,0.713095238)(38,0.81413690475)    };
\node [above] at (axis cs: 25,0.90520833325) {$A$};
\node [below] at (axis cs:  29,0.45744047600000004 ) {$B$};
\node [above] at (axis cs:  38,0.81413690475 ) {$C$};
\end{axis}
\end{tikzpicture}
\end{adjustbox}
\\(b) Impact on hit ratio on sudden drift in user behavior
\end{minipage}
\hfill
\begin{minipage}[t]{.24\linewidth}\centering
\begin{adjustbox}{width=\linewidth,totalheight=1.45in}
 \begin{tikzpicture} 
\begin{axis}[
    xlabel={Decay rate (k)},
    ylabel={Avg. Hit ratio},
    xmin=0.3, xmax=1.1,
    ymin=0.58, ymax=0.75,
    xtick={0.4,0.5,0.6,0.7,0.8,0.9,1.0},
    ytick={0.6,0.62,0.64,0.66,0.68,0.7,0.72},
    ymajorgrids=true,
    grid style=dashed,
]

\addplot[
    color=blue,
    mark=*,
    mark size=1pt
    ]
    coordinates {
 (0.4,0.662)(0.5,0.683) (0.6,0.705)(0.7,0.690)(0.8,0.679)(0.9,0.644) (1,0.629)};
\end{axis}
\end{tikzpicture}
\end{adjustbox}
\\(c) Hit ratio at different decay rates
\end{minipage}
\hfill
\begin{minipage}[t]{.24\linewidth}\centering
\begin{adjustbox}{width=\linewidth,totalheight=1.45in}
 \begin{tikzpicture} 
\begin{axis}[
    xlabel={Decay rate (k)},
    ylabel={Avg. Hit ratio},
    xmin=0.89, xmax=1.0,
    ymin=0.64, ymax=0.77,
    xtick={0.9,0.91,0.92,0.93,0.94,0.95,0.96,0.97,0.98,0.99},
    ytick={0.66,0.68,0.7,0.72,0.74,0.76},
    ymajorgrids=true,
    grid style=dashed,
]

\addplot[
    color=blue,
    mark=*,
    mark size=1pt
    ]
    coordinates {
(0.9,0.676597379)(0.91,0.686597379)(0.92,0.696597379)(0.93,0.698426647)(0.94,0.726228622)(0.95,0.718809922)(0.96,0.71881)(0.97,0.716169639)(0.98,0.704603152)(0.99,0.7000603152)};
\end{axis}
\end{tikzpicture}
\end{adjustbox}
\\(d)  Hit ratio at different cutoff values
\end{minipage}
\captionof{figure}{ Performance analysis of WIME under different conditions.}
\end{figure*}
\section{Experimental Results}
In this section, we evaluated our proposed algorithm in the following ways: 
\begin{enumerate}
\item How does our approach perform in different real world scenarios like \textit{user behavior drift} and \textit{ sequential usage} as discussed in the previous section?
\item How does it perform in comparison to state-of-the-art approaches on standard datasets?

\end{enumerate}

We prepared a dataset that captures user behaviour in real world scenarios. This has been used for in-app recommendation. We also explored various problem domains to identify which ones were analogous to ours. We concluded that category recommendation for the next personalized POI recommendations \cite{PRME} would be appropriate for comparison. Before recommending the next location the user is likely to visit, \cite{PRME} first predicts the category of the location that the user would be interested in visiting. This captures the essence of making recommendations based on user intent, much like what we propose. 

\subsection{Real world WIME dataset}
\subsubsection{Evaluation metrics}
To evaluate our approach on the dataset discussed in Section 3, we use hit ratio per day as an evaluation metric. For a given context $\displaystyle X_{t}$ at instance $\displaystyle t$, $\displaystyle I_{X_{t}}$ is the real user intent, we compare the intent $\displaystyle I_{X_{t}}$  against $\displaystyle I_{X_{t}^{*}}$ as predicted by our engine. Hit ratio for day $\displaystyle i$ is therefore defined as:
\begin{equation}
hit\ ratio_{i} =\frac{\#hits}{\#instances\ on\ day_{i}}
\end{equation}

\subsubsection{Observations}
Figure 5(a) illustrates how drifting average as discussed in Section 5.5 shows an improvement in the average hit ratio of all the users across days. Figure 5 (a) also displays how quickly WIME is able to start making reliable predictions. Point X was marked at day 6 while point Y was at day 12. Within a week the prediction engine achieved a hit ratio of nearly 81\%. This validates WIME's capability to quickly identify user intention patterns and make accurate predictions (\textbf {RQ3}). Figure 5(b) is a demonstration of how our method handles scenarios where the user shows a sudden shift in behavior pattern. For a particular user till point A, the user showed a regular behavior pattern, however beyond that, the user had a sudden change in one of the contextual factors (e.g: address change, work shift change, etc) which not only rendered the previously formed nodes obsolete, but also brought other changes in the daily activities both spatially and temporally. However, as we can observe that from point B onward, WIME adapts to the new pattern within a span of 2 weeks and reaches the previously obtained hit ratio at point C.

\subsubsection{Effect of decay rate}
Figure 5(c) depicts the impact of different decay rates on the performance of WIME. We attain the highest hit ratio at decay
 rate, k = 0.6, while the least when there is no decay (k=1.0)
 involved. These results validate the fact that maintaining node weights based on its relevance has an impact on the overall performance. 

\subsubsection{Effect of sequence comparison}
We evaluated our method without sequential information of the intents and compared it to the results provided when sequence was incorporated as a factor. We achieved a \textbf{4.3\% improvement} in the hit ratio when we considered sequence matching (\textbf {RQ2}), therefore confirming to the importance of sequences in intent prediction.

\subsubsection{Effect of prediction score cutoff}
As discussed in the step 3 of section 5.7, we filter nodes using a cutoff (c) for the sequence comparison step. We compared the hit ratio obtained for 0.90$\displaystyle{<}$c$\displaystyle{<}$0.99. As seen in Figure 5(d), we achieved the highest hit ratio at c = 0.94 and hence, the parameter was set to 0.94.

\subsubsection{Overall complexity}
We conducted experiments on mobile devices belonging to a diverse range of computational capabilities. The time complexity of prediction is roughly $\displaystyle{O(\log n)}$, $\displaystyle{n}$ being the number of nodes present at that instant. The time spent in the neighborhood-searching step is a major contributing factor in the runtime. Time added by the sequence comparison step is considered negligible as we make sequence comparisons for a fixed number of results (5 for our dataset) provided by the preceding step. The average prediction time achieved across all user devices was ~22-30 ms. The memory consumption is directly proportional to only the number of intention nodes in the vector space. We found that the average number of intention nodes retained for a user with decay rate of 0.6 is 198, taking upto 70kB of space on an average (\textbf {RQ1}).

\subsection{Foursquare}
We compare the performance of our approach with existing methods for category recommendation on user check-in data obtained from Foursquare \cite{foursquare} collected in the cities of Los Angeles and New York. The statistics of the dataset are summarized in Table 2.
\begin{table*}[t!]
        \centering
        
\begin{tabular}{|p{0.07\textwidth}|p{0.05\textwidth}|p{0.05\textwidth}|p{0.05\textwidth}|p{0.05\textwidth}|p{0.05\textwidth}|p{0.05\textwidth}|p{0.05\textwidth}|p{0.05\textwidth}|p{0.05\textwidth}|p{0.05\textwidth}|p{0.05\textwidth}|p{0.05\textwidth}|}
\hline 
 \multirow{2}{*}{Metrics} & \multicolumn{6}{l|}{
\ \ \ \ \ \ \ \ \ \ \ \ \ \ \ \ \ \ \ \
FourSquare-Los Angeles
} & \multicolumn{6}{l|}{
\ \ \ \ \ \ \ \ \ \ \ \ \ \ \ \ \ \ \ \
FourSquare-New York City
} \\
\cline{2-13} 
   & MF & FPMC & PRME & LBPR-N & Caser & WIME & MF & FPMC & PRME & LBPR-N & Caser & WIME \\
\hline 
 Prec@1 & 0.014 & 0.029 & 0.033 & 0.044 & \textbf{0.062} & 0.051 & 0.016 & 0.031 & 0.037 & 0.045 & \textbf{0.098} & 0.071 \\
\hline 
 Prec@5 & 0.051 & 0.082 & 0.074 & 0.172 & 0.243 & \textbf{0.245} & 0.059 & 0.083 & 0.080 & 0.174 & \textbf{0.354} & 0.352 \\
\hline 
 Prec@10 & 0.096 & 0.153 & 0.155 & 0.274 & 0.353 & \textbf{0.404} & 0.101 & 0.162 & 0.161 & 0.273 & 0.506 & \textbf{0.621} \\
 \hline
\end{tabular}
\caption{Performance comparison results between various recommender systems. WIME outperforms MF, FPMC, PRME and LBPR-N significantly. It shows comparable results against Caser, which is a complex convolutional neural network model.}
\end{table*}

\subsubsection{Evaluation metrics}
For evaluation purpose, we consider Precision@N as the comparison metric. It is defined it as:
\begin{equation}
Precision@N=\frac{1}{|U|}\sqrt{ _{u\in U}{\frac{|R_{u,N} \cap R^{*}|}{|R^{*} |}}}
\end{equation}
where $\displaystyle |U|$ is the number of users for user set $\displaystyle U$, $\displaystyle R_{u,N}$ is the set of top-N recommendations for user $\displaystyle u$ and $\displaystyle R^{*}$ is the set of ground truth. 
\subsubsection{Observations}
We compare the results obtained by WIME with those provided by the following approaches in Table 3. As can be observed, WIME beats several state of the art methods and shows results comparable to a deep learning based methdology(CASER):\\
\begin{itemize}
\item \textbf{MF}: Matrix Factorization \cite{MFSurvey} is one of the most commonly used approaches in the conventional recommender systems, which factorizes the user-item preference matrix.
\item \textbf{FPMC}: This method \cite{FPMC} is the state-of-the-art recommendation method which considers user preferences and their personalized Markov chain to provide next item recommendation.
\item \textbf{PRME}: This approach \cite{PRME} uses the Euclidean distance between current location and next location as well as the distance between user and next location. The combination of two distances is used to predict the next POI.
\item \textbf{LBPR-N}: This approach \cite{LBPR} uses Bayesian Personalized Ranking to recommend next POI based on the predicted next category.
\item \textbf{Caser}: This sequential recommendation method \cite{Caser} uses a convolutional sequence embedding to make Top-N personalized recommendations.
\end{itemize}
\section{Applications}
The WIME methodology devised in this paper is very versatile in nature. By modifying the context taken into consideration, we can determine which intent must be predicted and show the respective recommendations. This has been tested and verified on different applications using implicit relevance feedback. 
\subsection{Music recommendation}
\begin{figure}[h!]
\caption{Sample application for contextual music recommendation}
\includegraphics[scale=0.22]{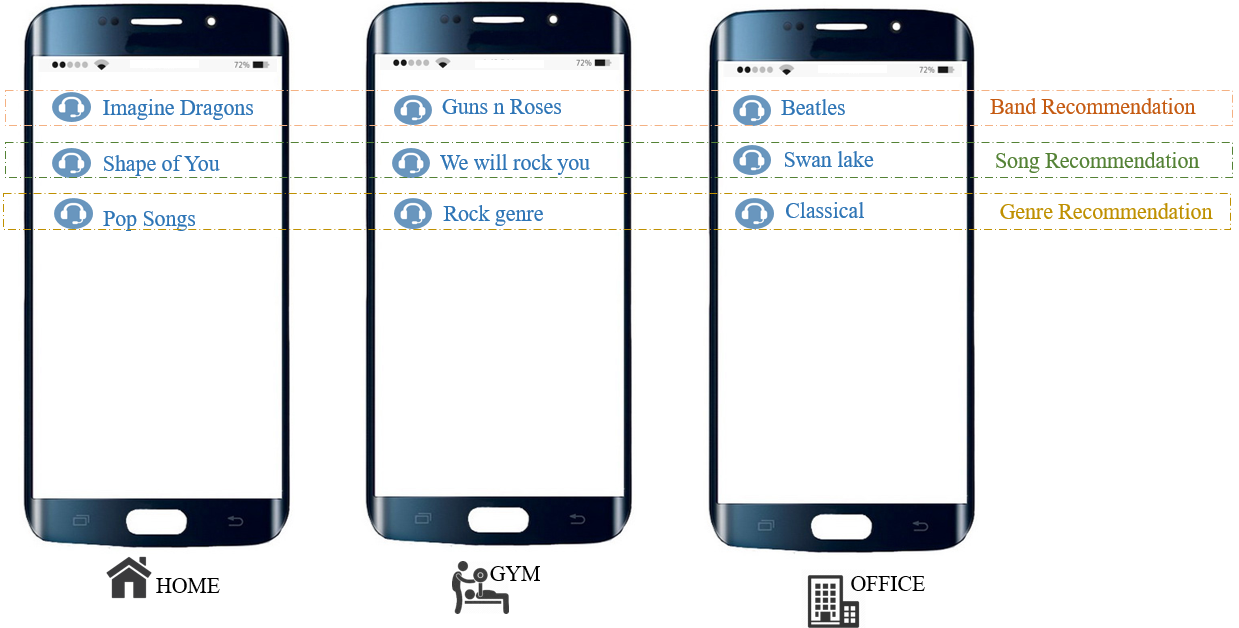}
\label{fig:appl1}
\end{figure}

The final intention to be predicted was changed to a subclass of music the user wants to listen to. We were able to generate recommendations in terms of song, band and genre recommendations by varying the granularity of the embeddings. An illustration of the sample application created is shown in Figure \ref{fig:appl1}.
\subsection{In-App prediction for launch time optimization}
\begin{figure}[h!]
\caption{Sample application for contextual in-app recommendation}
\includegraphics[scale=0.37]{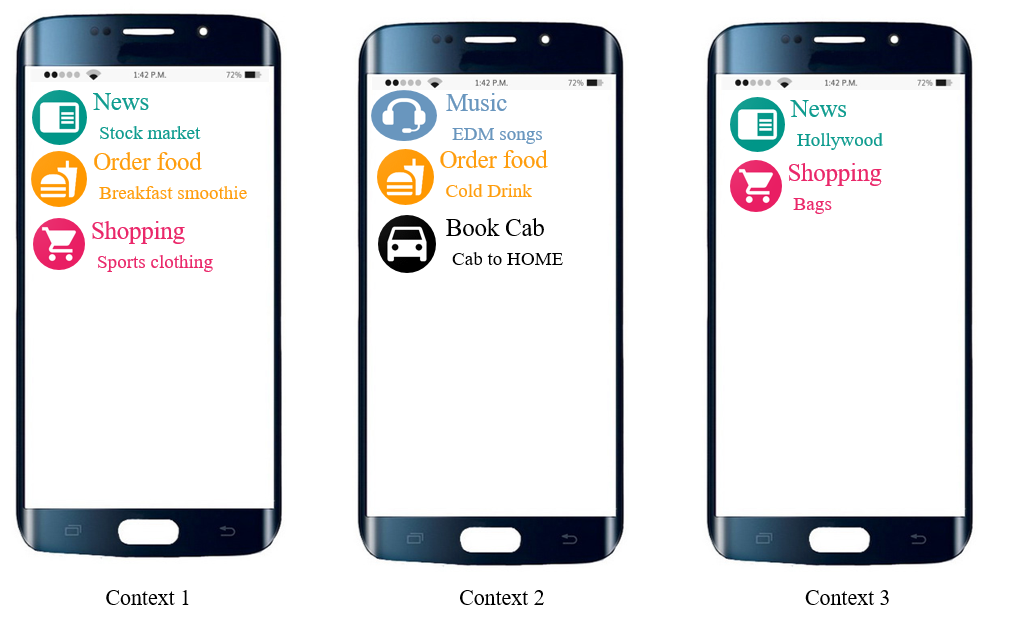}
\label{fig:appl2}
\end{figure}

The goal of this application was to reduce the amount of time user spends in navigating through various mobile applications and also preempt the loading of the next app screen so as to reduce app launch time. Here intentions are denoted as the action within an app that the user wants to take. Examples of these could be booking a cab for home at a particular context, ordering certain food items on a particular day of the week or booking a movie ticket every Friday. Instead of recommending just the next app, we go a step further and predict the action within the app too. Implicit feedback was collected for this to track the relevance of the feature in everyday use. We noted an average of 6\% decrease in navigation time spent per user. An illustration of the sample application created is shown in Figure \ref{fig:appl2}.

\section{Conclusion and Future Work}
In this paper, we address the problem of predicting user intent on device in real time and propose a novel method (WIME) for modelling user’s intents as weighted nodes in a multidimensional vector space in the form of embeddings. We handle the dynamic user behavior patterns such as gradual and sudden drifts in user action sequences without the need of complex deep learning methods. Experimental results on real world datasets indicate that WIME outperforms the state-of-the-art methods. For future work, we will encode more information into the embedding to capture a larger range of user context and improve performance of the proposed method by a large margin. WIME can also find applications on various devices with lower computational capabilities other than mobile devices.

\bibliographystyle{ACM-Reference-Format}
\bibliography{ref}


\begin{thebibliography}{21}


\ifx \showCODEN    \undefined \def \showCODEN     #1{\unskip}     \fi
\ifx \showDOI      \undefined \def \showDOI       #1{#1}\fi
\ifx \showISBNx    \undefined \def \showISBNx     #1{\unskip}     \fi
\ifx \showISBNxiii \undefined \def \showISBNxiii  #1{\unskip}     \fi
\ifx \showISSN     \undefined \def \showISSN      #1{\unskip}     \fi
\ifx \showLCCN     \undefined \def \showLCCN      #1{\unskip}     \fi
\ifx \shownote     \undefined \def \shownote      #1{#1}          \fi
\ifx \showarticletitle \undefined \def \showarticletitle #1{#1}   \fi
\ifx \showURL      \undefined \def \showURL       {\relax}        \fi
\providecommand\bibfield[2]{#2}
\providecommand\bibinfo[2]{#2}
\providecommand\natexlab[1]{#1}
\providecommand\showeprint[2][]{arXiv:#2}

\bibitem[\protect\citeauthoryear{Adomavicius, Sankaranarayanan, Sen, and
  Tuzhilin}{Adomavicius et~al\mbox{.}}{2005}]%
        {CARSEmergence}
\bibfield{author}{\bibinfo{person}{Gediminas Adomavicius},
  \bibinfo{person}{Ramesh Sankaranarayanan}, \bibinfo{person}{Shahana Sen},
  {and} \bibinfo{person}{Alexander Tuzhilin}.} \bibinfo{year}{2005}\natexlab{}.
\newblock \showarticletitle{Incorporating Contextual Information in Recommender
  Systems Using a Multidimensional Approach}.
\newblock \bibinfo{journal}{\emph{ACM Trans. Inf. Syst.}} \bibinfo{volume}{23},
  \bibinfo{number}{1} (\bibinfo{date}{Jan.} \bibinfo{year}{2005}),
  \bibinfo{pages}{103--145}.
\newblock
\showISSN{1046-8188}
\urldef\tempurl%
\url{https://doi.org/10.1145/1055709.1055714}
\showDOI{\tempurl}


\bibitem[\protect\citeauthoryear{Baeza-Yates, Jiang, Silvestri, and
  Harrison}{Baeza-Yates et~al\mbox{.}}{2015}]%
        {NextApp}
\bibfield{author}{\bibinfo{person}{Ricardo Baeza-Yates}, \bibinfo{person}{Di
  Jiang}, \bibinfo{person}{Fabrizio Silvestri}, {and} \bibinfo{person}{Beverly
  Harrison}.} \bibinfo{year}{2015}\natexlab{}.
\newblock \showarticletitle{Predicting The Next App That You Are Going To Use}.
  In \bibinfo{booktitle}{\emph{Proceedings of the Eighth ACM International
  Conference on Web Search and Data Mining}} \emph{(\bibinfo{series}{WSDM
  '15})}. \bibinfo{publisher}{ACM}, \bibinfo{address}{New York, NY, USA},
  \bibinfo{pages}{285--294}.
\newblock
\showISBNx{978-1-4503-3317-7}
\urldef\tempurl%
\url{https://doi.org/10.1145/2684822.2685302}
\showDOI{\tempurl}


\bibitem[\protect\citeauthoryear{Bentley}{Bentley}{1975}]%
        {kdTree}
\bibfield{author}{\bibinfo{person}{Jon~Louis Bentley}.}
  \bibinfo{year}{1975}\natexlab{}.
\newblock \showarticletitle{Multidimensional Binary Search Trees Used for
  Associative Searching}.
\newblock \bibinfo{journal}{\emph{Commun. ACM}} \bibinfo{volume}{18},
  \bibinfo{number}{9} (\bibinfo{date}{Sept.} \bibinfo{year}{1975}),
  \bibinfo{pages}{509--517}.
\newblock
\showISSN{0001-0782}
\urldef\tempurl%
\url{https://doi.org/10.1145/361002.361007}
\showDOI{\tempurl}


\bibitem[\protect\citeauthoryear{Binns, Lyngs, Van~Kleek, Zhao, Libert, and
  Shadbolt}{Binns et~al\mbox{.}}{2018}]%
        {ThirdPartyTracking}
\bibfield{author}{\bibinfo{person}{Reuben Binns}, \bibinfo{person}{Ulrik
  Lyngs}, \bibinfo{person}{Max Van~Kleek}, \bibinfo{person}{Jun Zhao},
  \bibinfo{person}{Timothy Libert}, {and} \bibinfo{person}{Nigel Shadbolt}.}
  \bibinfo{year}{2018}\natexlab{}.
\newblock \showarticletitle{Third Party Tracking in the Mobile Ecosystem}. In
  \bibinfo{booktitle}{\emph{Proceedings of the 10th ACM Conference on Web
  Science}} \emph{(\bibinfo{series}{WebSci '18})}. \bibinfo{publisher}{ACM},
  \bibinfo{address}{New York, NY, USA}, \bibinfo{pages}{23--31}.
\newblock
\showISBNx{978-1-4503-5563-6}
\urldef\tempurl%
\url{https://doi.org/10.1145/3201064.3201089}
\showDOI{\tempurl}


\bibitem[\protect\citeauthoryear{Christen}{Christen}{2006}]%
        {jaroComparison}
\bibfield{author}{\bibinfo{person}{Peter Christen}.}
  \bibinfo{year}{2006}\natexlab{}.
\newblock \showarticletitle{A Comparison of Personal Name Matching: Techniques
  and Practical Issues}. In \bibinfo{booktitle}{\emph{Proceedings of the Sixth
  IEEE International Conference on Data Mining - Workshops}}
  \emph{(\bibinfo{series}{ICDMW '06})}. \bibinfo{publisher}{IEEE Computer
  Society}, \bibinfo{address}{Washington, DC, USA}, \bibinfo{pages}{290--294}.
\newblock
\showISBNx{0-7695-2702-7}
\urldef\tempurl%
\url{https://doi.org/10.1109/ICDMW.2006.2}
\showDOI{\tempurl}


\bibitem[\protect\citeauthoryear{Feng, Li, Zeng, Cong, Chee, and Yuan}{Feng
  et~al\mbox{.}}{2015}]%
        {PRME}
\bibfield{author}{\bibinfo{person}{Shanshan Feng}, \bibinfo{person}{Xutao Li},
  \bibinfo{person}{Yifeng Zeng}, \bibinfo{person}{Gao Cong},
  \bibinfo{person}{Yeow~Meng Chee}, {and} \bibinfo{person}{Quan Yuan}.}
  \bibinfo{year}{2015}\natexlab{}.
\newblock \showarticletitle{Personalized Ranking Metric Embedding for Next New
  POI Recommendation}. In \bibinfo{booktitle}{\emph{Proceedings of the 24th
  International Conference on Artificial Intelligence}}
  \emph{(\bibinfo{series}{IJCAI'15})}. \bibinfo{publisher}{AAAI Press},
  \bibinfo{pages}{2069--2075}.
\newblock
\showISBNx{978-1-57735-738-4}
\urldef\tempurl%
\url{http://dl.acm.org/citation.cfm?id=2832415.2832536}
\showURL{%
\tempurl}


\bibitem[\protect\citeauthoryear{He, Li, and Liao}{He et~al\mbox{.}}{2017}]%
        {LBPR}
\bibfield{author}{\bibinfo{person}{Jing He}, \bibinfo{person}{Xin Li}, {and}
  \bibinfo{person}{Lejian Liao}.} \bibinfo{year}{2017}\natexlab{}.
\newblock \showarticletitle{Category-aware Next Point-of-interest
  Recommendation via Listwise Bayesian Personalized Ranking}. In
  \bibinfo{booktitle}{\emph{Proceedings of the 26th International Joint
  Conference on Artificial Intelligence}} \emph{(\bibinfo{series}{IJCAI'17})}.
  \bibinfo{publisher}{AAAI Press}, \bibinfo{pages}{1837--1843}.
\newblock
\showISBNx{978-0-9992411-0-3}
\urldef\tempurl%
\url{http://dl.acm.org/citation.cfm?id=3172077.3172143}
\showURL{%
\tempurl}


\bibitem[\protect\citeauthoryear{Knijnenburg and Berkovsky}{Knijnenburg and
  Berkovsky}{2017}]%
        {RecsysPrivacyTutorial}
\bibfield{author}{\bibinfo{person}{Bart~P. Knijnenburg} {and}
  \bibinfo{person}{Shlomo Berkovsky}.} \bibinfo{year}{2017}\natexlab{}.
\newblock \showarticletitle{Privacy for Recommender Systems: Tutorial
  Abstract}. In \bibinfo{booktitle}{\emph{Proceedings of the Eleventh ACM
  Conference on Recommender Systems}} \emph{(\bibinfo{series}{RecSys '17})}.
  \bibinfo{publisher}{ACM}, \bibinfo{address}{New York, NY, USA},
  \bibinfo{pages}{394--395}.
\newblock
\showISBNx{978-1-4503-4652-8}
\urldef\tempurl%
\url{https://doi.org/10.1145/3109859.3109935}
\showDOI{\tempurl}


\bibitem[\protect\citeauthoryear{Kobsa, Knijnenburg, and Livshits}{Kobsa
  et~al\mbox{.}}{2014}]%
        {clientBasedRecSys}
\bibfield{author}{\bibinfo{person}{Alfred Kobsa}, \bibinfo{person}{Bart~P.
  Knijnenburg}, {and} \bibinfo{person}{Benjamin Livshits}.}
  \bibinfo{year}{2014}\natexlab{}.
\newblock \showarticletitle{Let's Do It at My Place Instead?: Attitudinal and
  Behavioral Study of Privacy in Client-side Personalization}. In
  \bibinfo{booktitle}{\emph{Proceedings of the SIGCHI Conference on Human
  Factors in Computing Systems}} \emph{(\bibinfo{series}{CHI '14})}.
  \bibinfo{publisher}{ACM}, \bibinfo{address}{New York, NY, USA},
  \bibinfo{pages}{81--90}.
\newblock
\showISBNx{978-1-4503-2473-1}
\urldef\tempurl%
\url{https://doi.org/10.1145/2556288.2557102}
\showDOI{\tempurl}


\bibitem[\protect\citeauthoryear{Koren, Bell, and Volinsky}{Koren
  et~al\mbox{.}}{2009}]%
        {MFSurvey}
\bibfield{author}{\bibinfo{person}{Yehuda Koren}, \bibinfo{person}{Robert
  Bell}, {and} \bibinfo{person}{Chris Volinsky}.}
  \bibinfo{year}{2009}\natexlab{}.
\newblock \showarticletitle{Matrix Factorization Techniques for Recommender
  Systems}.
\newblock \bibinfo{journal}{\emph{Computer}} \bibinfo{volume}{42},
  \bibinfo{number}{8} (\bibinfo{date}{Aug.} \bibinfo{year}{2009}),
  \bibinfo{pages}{30--37}.
\newblock
\showISSN{0018-9162}
\urldef\tempurl%
\url{https://doi.org/10.1109/MC.2009.263}
\showDOI{\tempurl}


\bibitem[\protect\citeauthoryear{Liu, Pham, Cong, and Yuan}{Liu
  et~al\mbox{.}}{2017}]%
        {location-based-POI}
\bibfield{author}{\bibinfo{person}{Yiding Liu},
  \bibinfo{person}{Tuan-Anh~Nguyen Pham}, \bibinfo{person}{Gao Cong}, {and}
  \bibinfo{person}{Quan Yuan}.} \bibinfo{year}{2017}\natexlab{}.
\newblock \showarticletitle{An Experimental Evaluation of Point-of-interest
  Recommendation in Location-based Social Networks}.
\newblock \bibinfo{journal}{\emph{Proc. VLDB Endow.}} \bibinfo{volume}{10},
  \bibinfo{number}{10} (\bibinfo{date}{June} \bibinfo{year}{2017}),
  \bibinfo{pages}{1010--1021}.
\newblock
\showISSN{2150-8097}
\urldef\tempurl%
\url{https://doi.org/10.14778/3115404.3115407}
\showDOI{\tempurl}


\bibitem[\protect\citeauthoryear{Navarro}{Navarro}{2001}]%
        {levenshtein}
\bibfield{author}{\bibinfo{person}{Gonzalo Navarro}.}
  \bibinfo{year}{2001}\natexlab{}.
\newblock \showarticletitle{A Guided Tour to Approximate String Matching}.
\newblock \bibinfo{journal}{\emph{ACM Comput. Surv.}} \bibinfo{volume}{33},
  \bibinfo{number}{1} (\bibinfo{date}{March} \bibinfo{year}{2001}),
  \bibinfo{pages}{31--88}.
\newblock
\showISSN{0360-0300}
\urldef\tempurl%
\url{https://doi.org/10.1145/375360.375365}
\showDOI{\tempurl}


\bibitem[\protect\citeauthoryear{Quadrana, Cremonesi, and Jannach}{Quadrana
  et~al\mbox{.}}{2018}]%
        {SeqAwareSurvey}
\bibfield{author}{\bibinfo{person}{Massimo Quadrana}, \bibinfo{person}{Paolo
  Cremonesi}, {and} \bibinfo{person}{Dietmar Jannach}.}
  \bibinfo{year}{2018}\natexlab{}.
\newblock \showarticletitle{Sequence-Aware Recommender Systems}.
\newblock \bibinfo{journal}{\emph{ACM Comput. Surv.}} \bibinfo{volume}{51},
  \bibinfo{number}{4}, Article \bibinfo{articleno}{66} (\bibinfo{date}{July}
  \bibinfo{year}{2018}), \bibinfo{numpages}{36}~pages.
\newblock
\showISSN{0360-0300}
\urldef\tempurl%
\url{https://doi.org/10.1145/3190616}
\showDOI{\tempurl}


\bibitem[\protect\citeauthoryear{Rendle, Freudenthaler, and
  Schmidt-Thieme}{Rendle et~al\mbox{.}}{2010}]%
        {FPMC}
\bibfield{author}{\bibinfo{person}{Steffen Rendle}, \bibinfo{person}{Christoph
  Freudenthaler}, {and} \bibinfo{person}{Lars Schmidt-Thieme}.}
  \bibinfo{year}{2010}\natexlab{}.
\newblock \showarticletitle{Factorizing Personalized Markov Chains for
  Next-basket Recommendation}. In \bibinfo{booktitle}{\emph{Proceedings of the
  19th International Conference on World Wide Web}} \emph{(\bibinfo{series}{WWW
  '10})}. \bibinfo{publisher}{ACM}, \bibinfo{address}{New York, NY, USA},
  \bibinfo{pages}{811--820}.
\newblock
\showISBNx{978-1-60558-799-8}
\urldef\tempurl%
\url{https://doi.org/10.1145/1772690.1772773}
\showDOI{\tempurl}


\bibitem[\protect\citeauthoryear{Rodr\'{\i}guez-Hern\'{a}ndez and
  Ilarri}{Rodr\'{\i}guez-Hern\'{a}ndez and Ilarri}{2016}]%
        {CARSMobile2}
\bibfield{author}{\bibinfo{person}{Mar\'{\i}a del~Carmen
  Rodr\'{\i}guez-Hern\'{a}ndez} {and} \bibinfo{person}{Sergio Ilarri}.}
  \bibinfo{year}{2016}\natexlab{}.
\newblock \showarticletitle{Pull-based Recommendations in Mobile Environments}.
\newblock \bibinfo{journal}{\emph{Comput. Stand. Interfaces}}
  \bibinfo{volume}{44}, \bibinfo{number}{C} (\bibinfo{date}{Feb.}
  \bibinfo{year}{2016}), \bibinfo{pages}{185--204}.
\newblock
\showISSN{0920-5489}
\urldef\tempurl%
\url{https://doi.org/10.1016/j.csi.2015.08.002}
\showDOI{\tempurl}


\bibitem[\protect\citeauthoryear{Sassi, Mellouli, and Yahia}{Sassi
  et~al\mbox{.}}{2017}]%
        {CARSMobileSassi}
\bibfield{author}{\bibinfo{person}{Imen~Ben Sassi}, \bibinfo{person}{Sehl
  Mellouli}, {and} \bibinfo{person}{Sadok~Ben Yahia}.}
  \bibinfo{year}{2017}\natexlab{}.
\newblock \showarticletitle{Context-aware recommender systems in mobile
  environment: On the road of future research}.
\newblock \bibinfo{journal}{\emph{Inf. Syst.}}  \bibinfo{volume}{72}
  (\bibinfo{year}{2017}), \bibinfo{pages}{27--61}.
\newblock


\bibitem[\protect\citeauthoryear{Tang and Wang}{Tang and Wang}{2018}]%
        {Caser}
\bibfield{author}{\bibinfo{person}{Jiaxi Tang} {and} \bibinfo{person}{Ke
  Wang}.} \bibinfo{year}{2018}\natexlab{}.
\newblock \showarticletitle{Personalized Top-N Sequential Recommendation via
  Convolutional Sequence Embedding}. In \bibinfo{booktitle}{\emph{Proceedings
  of the Eleventh ACM International Conference on Web Search and Data Mining}}
  \emph{(\bibinfo{series}{WSDM '18})}. \bibinfo{publisher}{ACM},
  \bibinfo{address}{New York, NY, USA}, \bibinfo{pages}{565--573}.
\newblock
\showISBNx{978-1-4503-5581-0}
\urldef\tempurl%
\url{https://doi.org/10.1145/3159652.3159656}
\showDOI{\tempurl}


\bibitem[\protect\citeauthoryear{Winkler}{Winkler}{1990}]%
        {winkler}
\bibfield{author}{\bibinfo{person}{William~E. Winkler}.}
  \bibinfo{year}{1990}\natexlab{}.
\newblock \showarticletitle{String Comparator Metrics and Enhanced Decision
  Rules in the Fellegi-Sunter Model of Record Linkage}. In
  \bibinfo{booktitle}{\emph{Proceedings of the Section on Survey Research}}.
  \bibinfo{pages}{354--359}.
\newblock


\bibitem[\protect\citeauthoryear{Yin, Zhou, Cui, Wang, Zheng, and Hung}{Yin
  et~al\mbox{.}}{2016}]%
        {interestDriftPOI}
\bibfield{author}{\bibinfo{person}{Hongzhi Yin}, \bibinfo{person}{Xiaofang
  Zhou}, \bibinfo{person}{Bin Cui}, \bibinfo{person}{Hao Wang},
  \bibinfo{person}{Kai Zheng}, {and} \bibinfo{person}{Nguyen Quoc~Viet Hung}.}
  \bibinfo{year}{2016}\natexlab{}.
\newblock \showarticletitle{Adapting to User Interest Drift for {POI}
  Recommendation}.
\newblock \bibinfo{journal}{\emph{{IEEE} Trans. Knowl. Data Eng.}}
  \bibinfo{volume}{28}, \bibinfo{number}{10} (\bibinfo{year}{2016}),
  \bibinfo{pages}{2566--2581}.
\newblock


\bibitem[\protect\citeauthoryear{Yuan, Cong, and Sun}{Yuan
  et~al\mbox{.}}{2014}]%
        {foursquare}
\bibfield{author}{\bibinfo{person}{Quan Yuan}, \bibinfo{person}{Gao Cong},
  {and} \bibinfo{person}{Aixin Sun}.} \bibinfo{year}{2014}\natexlab{}.
\newblock \showarticletitle{Graph-based point-of-interest recommendation with
  geographical and temporal influences}. In
  \bibinfo{booktitle}{\emph{Proceedings of the 23rd ACM International
  Conference on Conference on Information and Knowledge Management}}. ACM,
  \bibinfo{pages}{659--668}.
\newblock


\bibitem[\protect\citeauthoryear{Zimdars, Chickering, and Meek}{Zimdars
  et~al\mbox{.}}{2001}]%
        {MC}
\bibfield{author}{\bibinfo{person}{Andrew Zimdars},
  \bibinfo{person}{David~Maxwell Chickering}, {and}
  \bibinfo{person}{Christopher Meek}.} \bibinfo{year}{2001}\natexlab{}.
\newblock \showarticletitle{Using Temporal Data for Making Recommendations}. In
  \bibinfo{booktitle}{\emph{Proceedings of the Seventeenth Conference on
  Uncertainty in Artificial Intelligence}} \emph{(\bibinfo{series}{UAI'01})}.
  \bibinfo{publisher}{Morgan Kaufmann Publishers Inc.}, \bibinfo{address}{San
  Francisco, CA, USA}, \bibinfo{pages}{580--588}.
\newblock
\showISBNx{1-55860-800-1}
\urldef\tempurl%
\url{http://dl.acm.org/citation.cfm?id=2074022.2074093}
\showURL{%
\tempurl}


\end{thebibliography}

\end{document}